\documentclass{article}

\usepackage{arxiv}

\usepackage[utf8]{inputenc} 
\usepackage[T1]{fontenc}    
\usepackage{hyperref}       
\usepackage{url}            
\usepackage{booktabs}       
\usepackage{amsfonts}       
\usepackage{nicefrac}       
\usepackage{microtype}      
\usepackage{lipsum}
\usepackage{soul}
\usepackage{color}
\usepackage{float}
\usepackage{graphicx}
\usepackage{caption}
\usepackage{subcaption}
\usepackage{titlesec}
\usepackage{multirow}
\usepackage{amsmath}
\usepackage{soul,color}
\usepackage[section]{placeins}
\usepackage{array, makecell} %
\usepackage{float}
\usepackage[section]{placeins}
\usepackage{wrapfig}
\usepackage{enumitem}
\usepackage{adjustbox}

\usepackage[nodisplayskipstretch]{setspace}

\title{Semi-supervised teacher-student deep neural network for materials discovery}
\author{
   \And
Daniel Gleaves\\
  Department of Computer Science and Engineering\\
  University of South Carolina\\
  Columbia, SC 29201 \\
  \And
Edirisuriya M. Dilanga Siriwardane, Yong Zhao\\
  Department of Computer Science and Engineering\\
  University of South Carolina\\
  Columbia, SC 29201 \\
  \And
 Nihang Fu\\
  Department of Computer Science and Engineering\\
  University of South Carolina\\
  Columbia, SC 29201 \\
  \And
 Jianjun Hu \thanks{Corresponding author: J.H. (http://www.cse.sc.edu/~jianjunh)}\\
  Department of Computer Science and Engineering\\
  University of South Carolina\\
  Columbia, SC, 29201, USA \\
  \texttt{jianjunh@cse.sc.edu}
}

\begin{document}
\maketitle

\begin{abstract}

Data driven generative machine learning models have recently emerged as one of the most promising approaches for new materials discovery. While the generator models can generate millions of candidates, it is critical to train fast and accurate machine learning models to filter out stable, synthesizable materials with desired properties. However, such efforts to build supervised regression or classification screening models have been severely hindered by the lack of unstable or unsynthesizable samples, which usually are not collected and deposited in materials databases such as ICSD and Materials Project (MP). At the same time, there are a significant amount of unlabelled data available in these databases. Here we propose a semi-supervised deep neural network (TSDNN) model for high-performance formation energy and synthesizability prediction, which is achieved via its unique teacher-student dual network architecture and its effective exploitation of the large amount of unlabeled data. For formation energy based stability screening, our semi-supervised classifier achieves an absolute 10.3\% accuracy improvement compared to the baseline CGCNN regression model. For synthesizability prediction, our model significantly increases the baseline PU learning's true positive rate from 87.9\% to 97.9\% using 1/49 model parameters.
 To further prove the effectiveness of our models, we combined our TSDNN-energy and TSDNN-synthesizability models with our CubicGAN generator to discover novel stable cubic structures. Out of 1000 recommended candidate samples by our models, 512 of them have negative formation energies as validated by our DFT formation energy calculations.  Our experimental results show that our semi-supervised deep neural networks can significantly improve the screening accuracy in large-scale generative materials design. 
\end{abstract}


\section{Introduction}


Machine learning based screening models have emerged as one of the most promising approaches for discovery of new materials either from repositories  of known materials \cite{sendek2017holistic,sendek2018machine,chen2020critical} or from hypothetical materials with compositions \cite{dan2020generative,song2021computational,song2020machine} or/and structures \cite{ren2020inverse,yong2021htd} generated by generative deep learning models or by crystal structure prediction algorithms \cite{oganov2019structure}. While existing materials repositories such as ICSD \cite{bergerhoff1987crystallographic} and Materials Project \cite{Jain2013} can be conveniently used for finding known synthesizable materials with new potential new functions, the success rate to discovering materials with extremely novel properties is severely constrained by the limited diversity and the number of known materials: the ICSD has only about \~ 200,000 crystal materials compared to the almost infinite chemical space. To search novel materials in uncharted chemical space, it is important to develop the capability to screen stable and synthesizable hypothetical materials \cite{aykol2019network,jang2020structure} out of the candidates generated by generative models or CSP algorithms and then apply high-performance materials property prediction models to find desired candidates \cite{omee2021scalable,louis2020graph}. 

Given a material's structure, its structural stability can be estimated by calculating its formation energy using first-principles computations such as density functional theory (DFT) and the phase
stability of a structure can be quantified by the energy above hull (Ehull) \cite{islam2020computational}. However, DFT based calculation of formation energy or Ehull is too computationally expensive, which leads to a large number of machine learning models for formation energy/enthalpy prediction \cite{huang2020practicing,peterson2021materials} based on composition without \cite{jha2018elemnet,jha2019enhancing, jha2021enabling,zhang2020machine,goodall2020predicting,peterson2021materials,krajewski2020extensible,bartel2020critical} or with structures \cite{xie2018crystal,chen2019graph,louis2020graph,fung2021benchmarking,jha2021enabling}. However despite the development of more than a dozen of formation energy/enthalpy prediction models, they all suffer from a neglected strong bias from the training data: most of the training samples from the repositories of know materials are stable structures with negative formation energy. For example, out of the 138,613 samples of Materials Project database, only 11,340 samples have positive formation energy. This makes it difficult to train good supervised classification or regression models that can differentiate stable materials from the unstable candidates. 


These methods usually formulate the formation energy prediction problem as a regression problem with models trained with a majority of negative formation energy. However, such formation energy prediction models are most interesting when they can be used to differentiate stable versus non-stable hypothetical materials, most of which tend to be unstable and have positive formation energy. Despite the claimed high accuracy of these models \cite{bartel2020critical,huang2020practicing}, they are mainly evaluated on the stable materials with negative formation energy, leading to their questionable extrapolation performance on out-of-distribution non-stable materials with positive formation energy \cite{xiong2020evaluating}. The question here is how can we train a ML models with a majority of samples with only negative formation energy while they are expected to differentiate stable materials with negative formation energy from unstable materials with positive formation energy. In addition to this issue, it is argued that the accurate prediction of formation alone does not correspond exactly to high accuracy of predicting stability which can be better measured by the quantity $\Delta$Hf and be obtained by a convex hull construction in formation enthalpy ($\Delta$Hf)-composition space \cite{bartel2020critical}. 

Synthesizability of a hypothetical material is another important property needed for effective materials screening \cite{szczypinski2021can,gao2020synthesizability} which is challenging to predict accurately \cite{davydov2019predicting}. It is found that many naive generative models for molecules tend to generate unsynthesizable candidates \cite{gao2020synthesizability}. Unfortunately, synthesizability is much more challenging to be predicted using ML models or other computational methods \cite{kovnir2021predictive,davydov2019predicting}. One approach is to predict the synthesis path given a material composition \cite{aykol2021rational,szymanski2021toward,aykol2021rational,malik2021predicting,shibukawa2020compret}; however, these approaches are newly emerging and cannot yet be applied to the large scale of hypothetical materials.  Another option is the ML based models for materials synthesizability prediction. For inorganic materials, a recent study using the Positive and Unlabelled semi-supervised machine learning algorithm (PU-learning) \cite{jang2020structure} has been applied to predict synthesizability with promising results.

Here we propose a semi-supervised learning (SSL) approach for the materials formation energy and synthesizability prediction problems by considering both the database bias that most samples are stable, synthesizable materials with negative formation energy and the model application scenarios for which we need to apply the models to differentiate stable and unstable hypothetical materials. 
Semi-supervised learning \cite{zhu2005semi,van2020survey} has been widely and successfully used in computer vision\cite{ren2020not} , natural language processing\cite{ouali2020overview}, medical diagnosis \cite{wang2020classification} to mainly address the scarce annotation data issue or just to improve the performance using unlabelled data. However, despite the well-known small data issue in materials ML problems, semi-supervised learning has  rarely been used in such problems except in a few studies  \cite{huo2019semi,jang2020structure,kunselman2020semi} for materials synthesis classification, microstructure classification, and synthesizability prediction. 

SSL algorithms are developed on several fundamental assumptions \cite{van2020survey} including (1) the smoothness assumption: two samples close to each other in the input space tend to have similar labels; (2) low-density assumption: the decision boundary should not pass through high-density areas in the input space; (3) manifold assumption: data points on the same low-dimensional manifold should have the same label. These assumptions can be interpreted as specific instances of the cluster assumption: similar points tend to belong to the same group/cluster. There are two main category of SSL algorithms including graph based transductive methods which focus on label propagation and inductive methods which aim to build a ML model f:x-->y by incorporating unlabelled data either in pre-processing steps, directly inside the loss function, or via a pseudo-labeling step. SSL algorithms have demonstrated strong performance especially in the deep learning framework \cite{ouali2020overview}.

In this work, we exploit a deep learning based SSL framework, the teacher-student deep neural networks (TSDNN) \cite{pham2021meta} to address the lack of negative samples in synthesizability prediction and formation energy prediction. TSDNN is characterized by a dual-network architecture with a teacher model trained using a supervised signal and an unsupervised feedback signal from the student to improve the teacher's pseudo-labeling. The teacher provides pseudo-labels for unlabeled data for the student to learn from. Unlike the previous positive-unlabeled SSL algorithm for synthesizability prediction, our TSDNN algorithm requires much fewer models to be trained to determine negative samples and train a final model while achieving 5.3\% higher prediction accuracy improving the positive rate from 86.7\% to 92.9\%  using the same performance evaluation. Extensive experiments on the formation energy classifiers also show that our TSDNN can screen negative formation energies with 7.5\% higher precision, 10.3\% higher F1 score, and 9.7\% higher accuracy than the CGCNN regression model.

Our contributions in this paper can be summarized as follows:

\begin{itemize}

    \item We identify the inherent dataset bias in formation energy and synthesizability prediction problems and propose to formulate both as semi-supervised classification problems. 
    \item We exploit a novel teacher-student dual network deep neural network model framework to achieve high-performance semi-supervised learning for both formation energy and synthesizability classification. Compared to the previous approaches, our models achieved >10\% performance improvement with much simpler model structures with 98\% fewer model sizes.
    \item We evaluate our algorithms on different data set configurations and demonstrate the effectiveness and advantage of SSL for both problems.
    \item We apply our TSDNN based formation energy and synthesizability SSL model for screening new materials from the hypothetical cubic crystal materials and identify a set of new stable materials as verified by DFT formation energy calculations.
\end{itemize}

\section{Methods}

\subsection{The framework for generative design of materials}
We follow a generation-and-screening approach for the discovery of novel materials: first, we use generative deep learning algorithms to generate hypothetical crystal structures in a high-throughput manner with millions of candidates \cite{yong2021htd}. The generated candidates will then be screened quickly using formation energy and synthesizability machine learning models. Finally, a set of top screened candidates will be verified by DFT based formation energy calculation and phonon dispersion verification. It should be noted that generative algorithms of materials compositions \cite{dan2020generative} can also be used here to first generate and screen out top compositions that are then fed to crystal structure prediction algorithms for structure determination and follow-up DFT validation.

In this work, we use our recently developed CubicGAN algorithm \cite{zhao2021high} to generate 10 million hypothetical ternary cubic crystal structures of three space groups (221,225, 216) which are reduced to 2.5 million unique candidate cubic structures. With such a high volume of candidates, how to find the stable and synthesizable ones is almost like finding the needle in the haystack. To address this challenge, we develop semi-supervised deep learning based classification models for identifying hypothetical materials candidates with negative formation energy and high synthesizability respectively.



\subsection{Semi-supervised learning based screening models using teacher-student deep neural networks (TSDNN)}

\begin{figure}[ht]
  \begin{subfigure}[b]{0.39\linewidth}
    \centering
    \includegraphics[width=\linewidth]{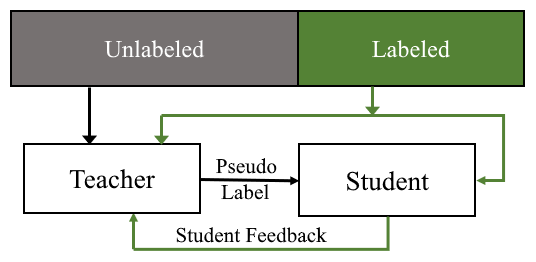}
    \caption{TSDNN Framework. $P_L$ and $N_L$ denote positive and negative labeled data respectively.}
    \label{fig:tsdnn_framework}
  \end{subfigure}
  \begin{subfigure}[b]{0.5\linewidth}
    \centering
    \includegraphics[width=\linewidth]{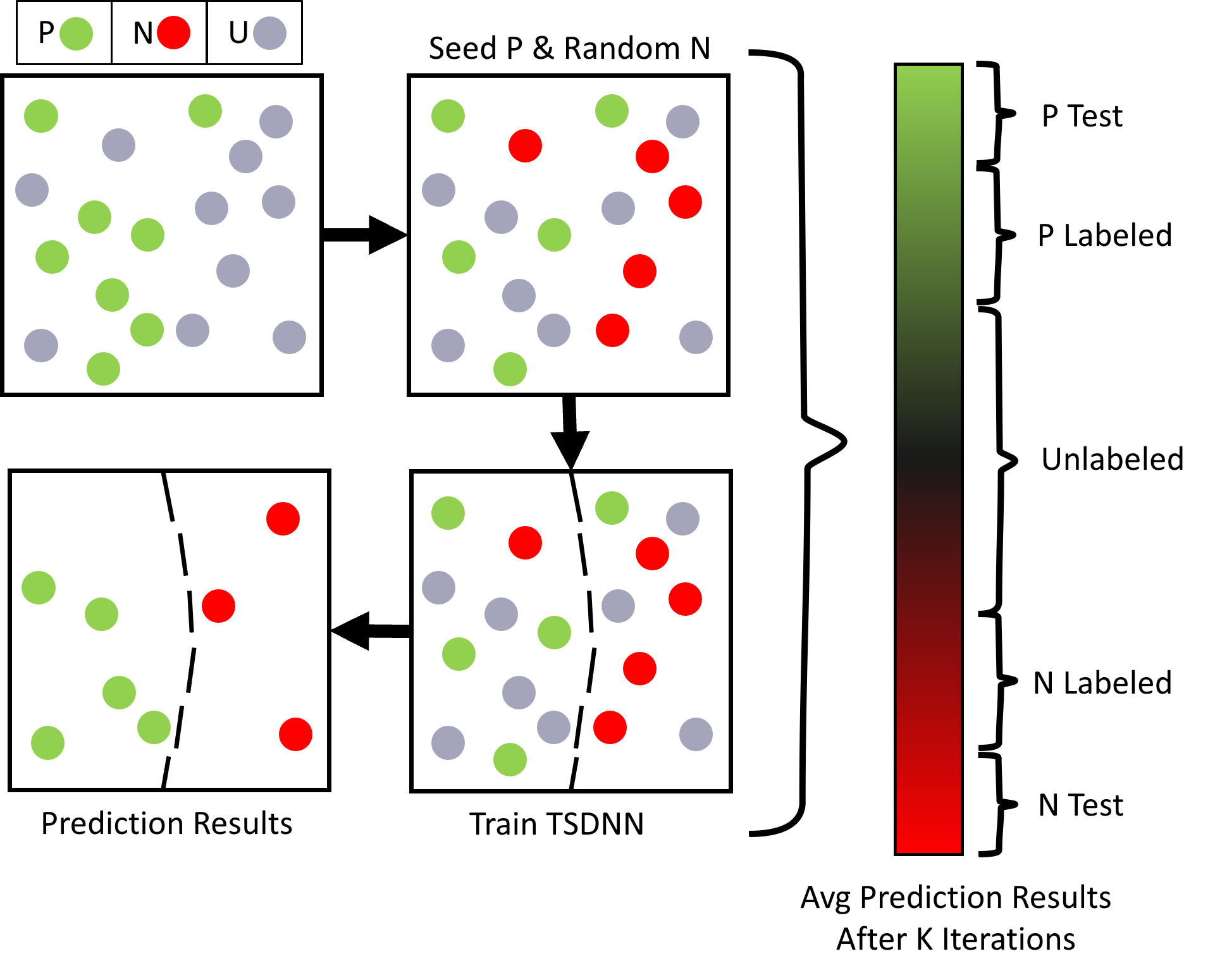}
    \caption{PU Learning-Based Synthesizability Dataset Generation Framework}
    \label{fig:pu_learning_framework}
  \end{subfigure}
  \caption{Dataset generation and TSDNN training framework and comparison to the PU-learning framework}
  \label{fig:framework}
\end{figure}

\begin{figure}[h]
  \centering
  \includegraphics[width=\linewidth]{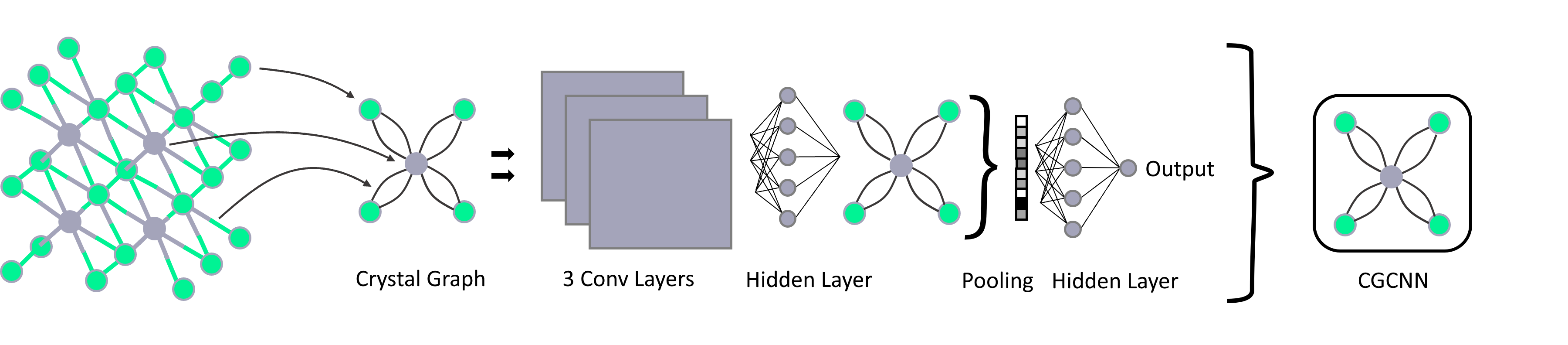}
  \caption{CGCNN architecture for structure based materials property prediction}
  \label{fig:cgcnn_architecture}
\end{figure}

In many problem domains and especially in materials science, the labeled dataset is too small, unbalanced, or has missing data classes. For example, there are only about ~2700 crystal materials with labelled thermal conductivity values \cite{gorai2016te} and less than 1700 annotated piezoelectric materials in the Materials Project (MP) database \cite{jain2013commentary}. Also in MP database, there are fewer than 8.2\% materials labelled with positive formation energy. 
As a result, it becomes difficult to train a well-converged supervised model to produce accurate classifications, especially on out-of-distribution samples. In this work, we propose to combine the TSDNN semi-supervised learning framework shown in Figure \ref{fig:framework} with a crystal graph convolutional neural network (CGCNN) \cite{xie2018crystal} (Figure \ref{fig:cgcnn_architecture}) for structure-based synthesizabiliy and formation energy prediction. The teacher-student deep neural network (TSDNN) leverages unlabeled data to overcome the issues with the too few labeled samples and the severe issue of lack of negative samples: unstable samples with positive formation energy or non-synthesizable samples.

TSDNN is a semi-supervised learning framework which is composed of two neural network models (Figure \ref{fig:tsdnn_framework}): a teacher network and a student network. These two models are trained in parallel. The teacher model generates pseudo labels on the unlabelled data which are then used to train the student network. The teacher model is trained with two objectives in our case: labeled data (synthesizability or formation energy classification) performance and a feedback signal \cite{pham2021meta} from the student model based on its performance on the labelled dataset. This feedback signal provides a guide for the teacher model in the case when the unlabeled samples are unlike the labeled data. The student model is trained only on unlabeled data with hard pseudo-labels provided by the teacher model. This leverages unlabeled data to improve further than supervised learning and smooth biases that may be found in the labeled data, such as through imbalances as with formation energy classification. 

The training process goes as follows: first, a batch of labeled and a batch of unlabeled data are sampled. The teacher's loss is calculated on the labeled batch. The teacher provides pseudo-labels for the student, with which the student will be updated. The student's performance on the labeled data is evaluated before the student is updated with the pseudo-labels and again after. The change in this performance that resulted from the teacher's pseudo-labels is used to calculate the student's feedback signal. This feedback signal is combined with the teacher's loss over labeled data to update the teacher model. This results in a student that is able to learn the true labels of a large set of unlabeled data without introducing biases from the labeled dataset.

Loss functions of our student and teacher network include:

\begin{equation}
\operatorname{L}^S_u=\mathbb{E}_{x_{u}}\left[\operatorname{CE}\left(T(x_u),S(x_u)\right)\right]
\end{equation}

Where $L_u$ represents the cross-entropy loss $CE$ on a batch of unlabeled dataset $X_u$ for the student $S$ network with respect to the labels produced by the teacher $T$. This is the student's only loss function.

\begin{equation}
\operatorname{L}^T_l=\mathbb{E}_{x_l,y_l}\left[\operatorname{CE}\left(y_l,T(x_l))\right)\right]
\end{equation}
Where $L^T_l$ represents the standard supervised cross-entropy loss $CE$ for a batch of labeled data $(x_l, y_l)$ for the teacher model $T$.

A feedback signal from the student model\cite{pham2021meta} is additionally included to further optimize the teacher model by improving its pseudo-labeling. This further reduces data bias introduction by introducing a dynamic teacher. While a static teacher model would replicate the implicit biases, this dynamic teacher is able to adapt, which leads to a less biased student model.

For the method to work efficiently, a few conditions must be met.
\begin{enumerate}[topsep=3pt,partopsep=3pt,itemsep=3pt,parsep=3pt]
    \item The labeled dataset should be accurate. Otherwise, the student model will be trying to optimize based on inaccurate teacher. The student feedback signal allows for some teacher inaccuracy to be resolved, but the majority of labels should be accurate.
    \item There should be a set of samples similar to the labeled dataset, or a subset therein, held within the unlabeled dataset. This provides an accurate foundation for the student that the teacher will be able to easily correctly classify. 
    \item There should be a set of representative samples from each class within the unlabeled dataset.
\end{enumerate}

To satisfy these criteria, we have introduced a novel approach of semi-supervised seeding similar to those found in semi-supervised clustering. By introducing a random distribution of data from the labeled set into the unlabeled set, we are able to effectively ensure that the student model has representative samples from each class that are similar to those found in the labeled set. 

In the teacher-student deep neural network framework, before training can commence, the dataset must be prepared for our semi-supervised framework. In the case of synthesizability, there is only positive data, so we must first identify candidate negative samples. This method is described in Section \ref{sec:synth-methods}. The labeled dataset should be optimized for accuracy by the smoothness assumption and low-density assumption. For both use-cases explored in this work, the labeled data selection is non-trivial. For synthesizability, selecting the most optimal negative labels, while resulting in fewer negative samples overall, results in the most optimal smoothness. For formation energy classification, the low-density assumption is the most significant to account for due to the large number of materials with near-zero formation energies, as shown in Figure \ref{fig:distributionEform}. Once the dataset is prepared and the SSL assumptions are provided for, the model is trained. 

\paragraph{Comparison with the PU-learning SSL framework:} 

Material synthesizability prediction is a positive and unknown (PU) learning problem, meaning that there are only materials with ICSD entries which have been previously synthesize and experimental materials which may or may not be able to be synthesized. For material synthesizability classification, we refer to one PU learning framework in particular \cite{jang2020structure} (Figure \ref{fig:pu_learning_framework}), which we use to generate an experimental labeled dataset and then optimize using our framework.

This PU learning framework is a modified transductive bagging support vector machine \cite{mordelet2014bagging}. In this framework, a model is trained with a random selection of unlabeled data set as the negative class of equal size to the positive class. This model then produces predictions on the remaining unlabeled set. After a given number of iterations, the unlabeled scores are averaged, resulting in a final score. The motivation is to identify a cluster of samples that lie apart from the positive class and any that lie in the middle should have an uncertain prediction score close to 0.5. Though there are results from a seemingly well-converged model, it will be unclear what prediction score threshold is reliable for true negative labels. For this reason, any supervised model trained from these results would be similarly unreliable. 

Using our semi-supervised framework, we were able to both identify the threshold of true negative samples and optimize the uncertain predictions close to 0.5. We optimize the materials with uncertain results by introducing the student feedback signal for the teacher, which results in a dynamic teacher that will be adjusted by the signal to generate pseudo-labels that coincide with the labeled data. This is integral for correctly classifying outlier samples, as the teacher will produce a pseudo label that, if the student then performs worse, will be adjusted. This allows the student model to be more robust and more reliable across all samples, as any out-of-distribution samples will be learned and optimized for performance on the labeled dataset. 

Simply using a semi-supervised pseudo-labeling model is not able to overcome the unreliability of the selected negative labels, as any inaccurate labels would be carried through the pseudo-labels. However, since our student model is trained only on unlabeled data and the student's predictions are optimized for labeled performance, as discussed above, we were able to identify the threshold at which the results became unreliable. Since the negative materials in the test set were withheld as a group instead of a random distribution, the performance of the model on these samples indicate the reliability of our data. It is not immediately clear whether this is from the labeled data or unlabeled data. However, by iteratively lowering the threshold of samples chosen as the negative set, we were able to increase the true negative rate performance until it became balanced with true positive rate, which indicated that we reached the reliability threshold. If the performance does not improve, this would indicate that the issue is with the unlabeled set and may benefit from semi-supervised seeding as discussed above.

The most significant difference between these two frameworks is that the PU learning framework is specialized for positive and unknown learning scenarios. Our framework provides a more robust and generalized approach that could be applied and modified to fit any scenario. With it, we were able to approach two distinct non-standard scenarios in which there is not a well-defined labeled dataset. For synthesizability, this is in the lack of a positive class and for formation energy it is that it is continuous data, lacking clear classification boundaries. In both scenarios, we were able to leverage our SSL framework to gain insight into the underlying motifs of the data and optimize our dataset formation.

\paragraph{CGCNN model for structure based classification:} The TSDNN model is a wrapper framework for semi-supervised classification, which can be combined with any material property prediction model. Here we adopt the CGCNN model for structure based synthesizablity and formation energy prediction. The CGCNN (Figure \ref{fig:cgcnn_architecture}) works by converting material structures in their unit cell into crystal graphs by encoding atoms as nodes and bonds as the edges between them. Encoding both the atomic features and bond interactions between atoms, the inherent structural characteristics can be learned. A convolutional neural network is then build on top of the crystal graphs to extract the feature representations to use for classification. 

\subsection{Synthesizability screening model using semi-supervised learning models} 
\label{sec:synth-methods}

Our synthesizability screening model (TSDNN-syn) is a binary classification model trained using the above-mentioned semi-supervised teacher-student neural network (TSDNN). 
We obtained the training dataset from the Materials Project database. More specifically, we obtained 125,619 materials with 48,146 of them being ICSD entries. As such, they  are labeled as 1 to indicate that they are synthesizable materials. However, in the case of synthesizability prediction, there are no known un-synthesizable materials to label as 0. There are only ICSD entries and virtual materials, the latter with an unknown synthesizability status. This lack of negative samples prevents a traditional supervised classification model from being trained as it normally would. To overcome this, we used a positive and unknown (PU) learning method \cite{jang2020structure} to identify materials with a low synthesizability score, as discussed above for the initial experimental dataset. In this framework, shown in Figure \ref{fig:tsdnn_framework}, our TSDNN model is trained 5 independent iterations. Each iteration, a random subset of the positive data is selected as seed data to be placed in the unlabeled set. A random subset of unlabeled dataset is selected, prior to adding the seed data, to be the negative set with a size equal to the number of samples in the positive set after removing the seed data. A TSDNN model is then trained on this data. This model makes predictions on the unlabeled samples not selected as the negative set. The final predicted scores are averaged across the 5 iterations to provide well converged results. We then selected the 48,146 lowest-scored materials (all below 0.33) to match the 48,146 positive samples. This provides a full labeled dataset with negative labels selected with more confidence than random selection to use as a foundation. 

This dataset could be directly used to trained a supervised or semi-supervised model, which was performed with the Balanced TSDNN and Supervised CGCNN models. However, since the negative labeled materials are selected as the result of an imperfect model's predictions, there will be false negatives introduced into the training data. This increases as materials are selected that had prediction scores closer to 0.5 than to 0.0. As a countermeasure to this, we leverage our semi-supervised model to gain insight into the dataset and select optimal negative samples. When trained with our semi-supervised model, the true negative rate is especially low compared to the true positive rate. However, when the threshold for negative samples is moved lower from 0.33, this performance improves. By utilizing this, we were able to determine the optimal negative class threshold to balance the true positive rate and true negative rate, which resulted in the improved performance of the Unbalanced TSDNN.

Random TSDNN: The Random TSDNN model uses a labeled dataset composed of the 48146 ICSD entry materials and an equal number of the lowest classification score samples from the PU learning dataset generation.

Unbalanced TSDNN: The Unbalanced TSDNN uses the 48146 ICSD entries with 9629 removed for the test set, resulting in 38517 total positive samples. The negative set is composed of the 6648 remaining negative samples that were below the reliability threshold after removing the 9629 for the test set.

TSDNN: The TSDNN model uses a training set of 48146 positive samples and 48146 negative samples, each with 9629 removed resulting in a training set of 38517 positive and 38517 negative samples. 

Supervised: The Supervised model uses the same labeled dataset as TSDNN.

The hyper-parameters of our TSDNN-syn model trainings are set as follows: 

\begin{table}[ht]
\centering
\caption{ Hyper-parameters for TSDNN training}

\begin{tabular}{|c|c|}
\hline
Hyperparameters              & Value \\ \hline
Epochs                       & 100   \\ \hline
Learning Rate                & 0.001 \\ \hline
Momentum                     & 0.9   \\ \hline
Weight Decay                 & 0     \\ \hline
Atomic Feature Length        & 90    \\ \hline
Hidden Feature Length        & 180   \\ \hline
Number of Convolution Layers & 3     \\ \hline
Number of Hidden Layers      & 1     \\ \hline
\end{tabular}
\label{tab:hyperparameters}
\end{table}

\subsection{Formation energy based screener using semi-supervised TSDNN framework} \label{fe-methods}

We designed two different TSDNN models for formation energy prediction to overcome biases inherent with previous methods due to having few samples with positive formation energy. We designed the first model, Separated TSDNN, to classify whether the formation energy of a material is above or below a threshold of -2.0 eV. We chose this threshold since there are many materials with slightly negative formation energy (-2.0, 0) that may be very structurally similar to those with slightly positive formation energies (0, 1.0). For this model, we used the materials with formation energies below -2.0 eV (n=5549) as positive samples in the labeled dataset. We selected an equal number of samples with the highest formation energies as negative samples. 

For the second model, Unseparated TSDNN, we used only materials with positive formation energies (n=2444) as negative samples and an equal number of randomly selected materials with negative formation energy as positive samples. This is optimized for a representative distribution of positive samples, with the intent of ensuring dataset smoothness and a low-density. This allows for improved smoothness by including samples with near-zero eV formation energies while still ensuring a low-density near the classification threshold of 0.0 eV. This is a general screener for positive vs. negative formation energy screening as opposed to the first approach, which is optimized for strictly low eV classification. This approach resulted in a high-precision model, where 78.4\% of samples with predicted scores greater than 0.5 have a formation energy of less than -2.0 eV and 99.0\% having a negative formation energy. It correctly classified 57.8\% of the possible samples with formation energies less than -2.0 eV. 

In both models, we use an unlabeled dataset with 500,000 CubicGAN-generated structures. These two models ensure there is a low sample density at the classification threshold. To use the dataset as-is with a threshold of 0.0 eV would result in a very high-density of materials at the threshold. As such, we use the different thresholds and data-selection methods to account for this. Each model has distinct benefits that are best suited for different applications, as discussed in Section \ref{sec:fe_perfomance}.

\begin{figure}[ht]
  \centering
  \begin{subfigure}{0.49\textwidth}
      \includegraphics[width=\textwidth]{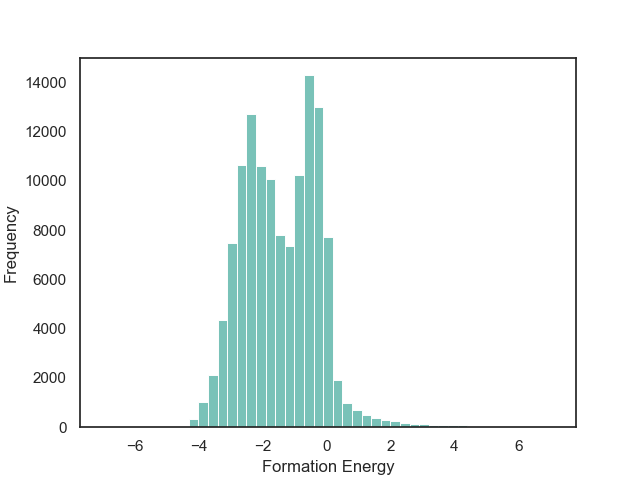}
      \caption{Material Project Formation Energies [eV/atom]}
      \label{fig:mp_fe_hist}
  \end{subfigure}
  \begin{subfigure}{0.5\textwidth}
      \includegraphics[width=\textwidth]{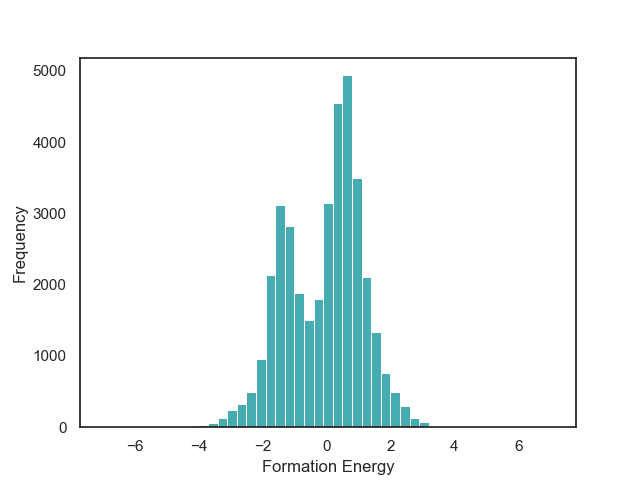}
      \caption{CubicGAN Test Set Formation Energies [eV/atom]}
      \label{fig:cubicgan_fe_hist}
  \end{subfigure}
  \caption{Distribution of formation energy for MP dataset and the Cubic test dataset.}
  \label{fig:distributionEform}
\end{figure}

We structure our datasets in this way to correct for biases and inconsistencies that models are ingrained with due to the unbalanced nature of formation energy datasets. As shown in Figure \ref{fig:mp_fe_hist}, the Material Project has an overwhelming majority of <0 eV materials. If trained from the raw data, it is likely that a model will bias heavily toward predicting >0 eV materials as being <0 eV. For this reason, we seek to combine the benefit of our TSDNN model with a balanced dataset to remove this bias. It is of particular importance that the model be unbiased when used with generated materials, such as those produced by our CubicGAN, as they contain many more >0 eV materials. We seek to apply our method to provide superior screening performance in identifying low formation energy materials.

\subsection{Evaluation criteria}

We evaluate the TSDNN-syn models based on their true positive rate on each model's respective test set. We use a prediction score boundary of 0.5 to determine a positive or negative sample classification. This classification performance can be expressed as

\begin{equation}
\operatorname{TPR}=\frac{TP}{TP + FN}
\end{equation}

where $TP$ is the number of true positive samples with predicted scores >= 0.5 and $FN$ is the number of true positive samples falsely negatively classified with a predicted score < 0.5. Since only positive samples are known, the true positive rate is the best indicator of performance in showing a model that accurately classifies true positive samples.

We evaluate the TSDNN-fe models on three metrics with variable formation energy thresholds: accuracy,  precision, and F1 Score. We again use a prediction score boundary of 0.5 to determine a positive or negative sample classification. The accuracy metric is shown as

\begin{equation}
    \operatorname{ACC(T)}=\frac{TP + TN}{(TP + FN) + (TN + FP)}
\end{equation}

where $TP$ denotes the number of samples with a formation energy below the threshold $T$ with predicted scores >= 0.5. $TN$ is the number of samples with a formation energy above $T$ with predicted scores < 0.5. $FN$ and $FP$ are the number false negative and false positive classifications using the same thresholds. 

The precision and recall metrics can be expressed as

\begin{equation}
    \operatorname{PR(T)}=\frac{TP}{TP + FP}
\end{equation}

\begin{equation}
    \operatorname{F1(T)}=\frac{2*P*R}{P + R}
\end{equation}

where $P$ is the model's precision and $R$ is the model's recall both with respect to the given formation energy threshold $T$.

\section{Experimental Results}
\label{sec:headings}

\subsection{Datasets}

We use inorganic material structures obtained from the Materials Project\cite{Jain2013,Ong2012b,Ong_2015} (MP) database for both our synthesizability prediction model and our formation energy prediction models. The MP database is a widely used material database consisting of materials obtained from the ICSD\cite{bergerhoff1987crystallographic} database or through high-throughput DFT calculations. In the case of synthesizability, we use the MP materials with ICSD entries as the positive dataset and the negative labels selected from the virtual MP materials as described in Section \ref{sec:synth-methods}. For our formation energy model, we use a combination of the MP database and a custom dataset of material structures generated by our CubicGAN model\cite{yong2021htd}. Our criteria for selecting positive and negative samples are detailed in Section \ref{sec:fe_perfomance}. Table \ref{tab:datasets} shows the source and number of samples in each dataset for each model. To compare the performance of our TSDNN models with the baseline PU-learning method, we first prepare a random test dataset in the same way as done in previous work \cite{jang2020structure}, which are composed of a random selection of 9629 positive (synthesizable) samples from the labeled set. We find our algorithm achieves 97.90\% true positive rate due to the test set materials being very structurally similar to those found in the training set. To validate that our model is able to accurately classify materials structurally different than those in the training set, we prepared a balanced test set composed of the 9629 negative samples with the lowest classification score from the PU learning dataset generation and a group of randomly selected 9629 positive samples. By introducing the negative samples, we are able to ensure that the model does not simply predict all materials as positive and has actually learned the structure features linked to synthesizability.



\begin{table}[H]
\caption{Training datasets}

\centering
\begin{adjustbox}{width=\textwidth}
\begin{tabular}{|l|l|l|l|l|l|l|l|l|l|}
\hline
\multicolumn{5}{|c|}{Synthesizability}      & \multicolumn{5}{c|}{Stability (formation energy)}                \\ \hline
\multicolumn{1}{|c|}{Model} &
  \multicolumn{1}{c|}{Labeled} &
  Src &
  \multicolumn{1}{c|}{Unlabeled} &
  Src &
  \multicolumn{1}{c|}{Model} &
  \multicolumn{1}{c|}{Labeled} &
  Src &
  \multicolumn{1}{c|}{Unlabeled} &
  Src \\ \hline
Supervised CGCNN      & 77035 & MP & 0     & N/A & Unseparated FE TSDNN & 4888  & MP & 500000 & CG     \\ \hline
Balanced TSDNN            & 77035 & MP & 29327 & MP  & Separated FE TSDNN      & 11078 & MP & 500000 & CG     \\ \hline
Unbalanced TSDNN & 45165 & MP & 29327 & MP  & CGCNN Regressor            & 20614 & MP & 0      & N/A  \\ \hline
PU-learning\cite{jang2020structure}  & 46781 & MP & 78734     & MP &              &  &  &       &  \\ \hline
\end{tabular}
\end{adjustbox}
\label{tab:datasets}
\end{table}

The Supervised CGCNN and Balanced TSDNN models use the same labeled datasets. The Balanced TSDNN model is trained using the remaining samples as the unlabeled set. This uses the unoptimized dataset provided from the dataset generation step. The Unbalanced TSDNN uses the optimized labeled dataset from the optimization step discussed in Section \ref{sec:synth-methods}.

Due to the fact that our CubicGAN generative model producing strictly cubic structures, we utilized only cubic Materials Project structures to train a formation-energy classification model to predict samples with negative formation energies. To achieve this, we used only the Material Project database's cubic structures to train our models. We used two selections of data for our formation energy models. The first model, the Distributed TSDNN, only materials with formation energies lower than 0.0 eV are used as negative data (n=2444). We then randomly selected an equal number from the remaining samples as positive data (n=2444). This allowed for a balanced labeled dataset with a solid distribution of negative formation energy samples represented. The second model, the Separated TSDNN, is trained using the lowest 25\% eV samples (n=5539) as positive data and the highest eV materials (n=5539) as negative data. This excludes the range of materials close to 0.0 eV. This motivation for this is to further separate the positive and negative classes in the input space. The CGCNN regression model is trained using the full cubic training dataset. We validate our formation energy models' performance by testing it on our own dataset of cubic structures produced by the CubicGAN with DFT-calculated formation energies. For each model, we used a test set of 36,847 CubicGAN-generated structures with DFT-calculated formation energies. This test set has 16,407 negative formation energy samples and 20,440 positive formation energy samples.

\subsection{Performance evaluation of TSDNN based semi-supervised learning}

We compare our TSDNN-syn and TSDNN-fe models against previous structure-based methods for predicting synthesizability and formation energies respectively. For synthesizability classification, we compare against the previous semi-supervised method of PU learning \cite{jang2020structure}. In the case of formation energy screening, we compare against a CGCNN regression model. We perform additional performance validation of our method by screening 2,545,713 novel CubicGAN-generated materials and selecting the top 1,000 for analysis. We perform DFT calculations to calculate their formation energies to analyze their stability and likely synthesizability.

\subsubsection{Synthesizability classification performance}

Due to the lack of known true negative samples (non-synthesizable samples) for synthesizability prediction, true positive rate is used here to evaluate the performance of the synthesizability prediction models. We include the accuracy metric for our tests as we utilize our method for selecting high-quality negative samples in addition to true positive rate. This is to validate that there is not simply a positive bias that results in a high true positive rate and there is in fact an observable differentiation in the model predictions. This is not possible with the previous method.

\begin{table}[H]
\centering
\caption{Synthesizability Results Comparison }
\begin{tabular}{|l|r|r|l|}
\hline
\multicolumn{1}{|c|}{Model} & \multicolumn{1}{c|}{TPR} & \multicolumn{1}{c|}{Accuracy} & \multicolumn{1}{c|}{Test Set} \\ \hline
Supervised CGCNN (baseline)                & 81.60\%       &  62.73\%         & 9629 Holdout    \\ \hline
Balanced TSDNN (ours)                     & 81.20\%   &  56.40\%               & 9629 Holdout                  \\ \hline
Seeded TSDNN (ours)               & 93.80\%      & 91.48\%             & 9629 Unlabeled                \\ \hline
Unbalanced TSDNN (ours)           & 92.90\%      &  \textbf{94.11\%}           & 9629 Holdout                  \\ \hline
Balanced TSDNN* (ours)              & \textbf{97.90\%}   &   N/A            & 9629 Holdout                  \\ \hline
PU-learning (baseline)\cite{jang2020structure}*           & 87.90\%    & N/A              & 9629 Holdout                  \\ \hline
\end{tabular}
\label{tab:synth-results}
\end{table}


We show the results of our synthesizability prediction in Table \ref{tab:synth-results}. The results denoted with a * were evaluated using a random subset of positive materials as the test set only. This is the most direct comparison to the PU-learning model\cite{jang2020structure}, which uses this evaluation method. We first tested our Balanced TSDNN model using this test set, and its true positive rate (TPR) performance was particularly high, with a TPR of 97.90\%, compared to the PU learning model that had a TPR of 87.90\%. However, this test did not have negative samples in the test set. As such, there was no way to ensure this high performance was not a result of a model bias. For this reason, we used the full test dataset (P Test + N Test) shown in Figure \ref{fig:pu_learning_framework}. This allows us to evaluate both the TPR and accuracy metrics. The Balanced TSDNN was trained using the full labeled dataset and a small unlabeled dataset to compare to the strictly supervised CGCNN classifier method. These two models have equivalent performance, with the Supervised CGCNN achieving an 81.60\% TPR and the Balanced TSDNN achieving an 81.20\% TPR. To improve on this and benefit from semi-supervised learning, we then use the optimized dataset described in Section \ref{sec:synth-methods} for training the Unbalanced TSDNN model, which achieved the highest accuracy of 94.11\% along with TPR of 92.90\%. We also evaluate this model by moving the test data into the unlabeled dataset for the Seeded TSDNN test. We use this test to evaluate the pseudo-labeling ability of our teacher model and to show that the true labels of data in the unlabeled set are learned correctly. The Seeded TSDNN achieves a TPR of 92.90\% and accuracy of 91.48\%, which demonstrates accurate teacher pseudo-labelling for unlabeled data. It increased the TPR of the Unbalanced TSDNN from 92.90\% to 93.80\%. This is the best comparison to real-world performance, as the unlabeled data would be the desired data to be classified.

\begin{figure}[ht]
  \centering
  \includegraphics[width=0.49\linewidth]{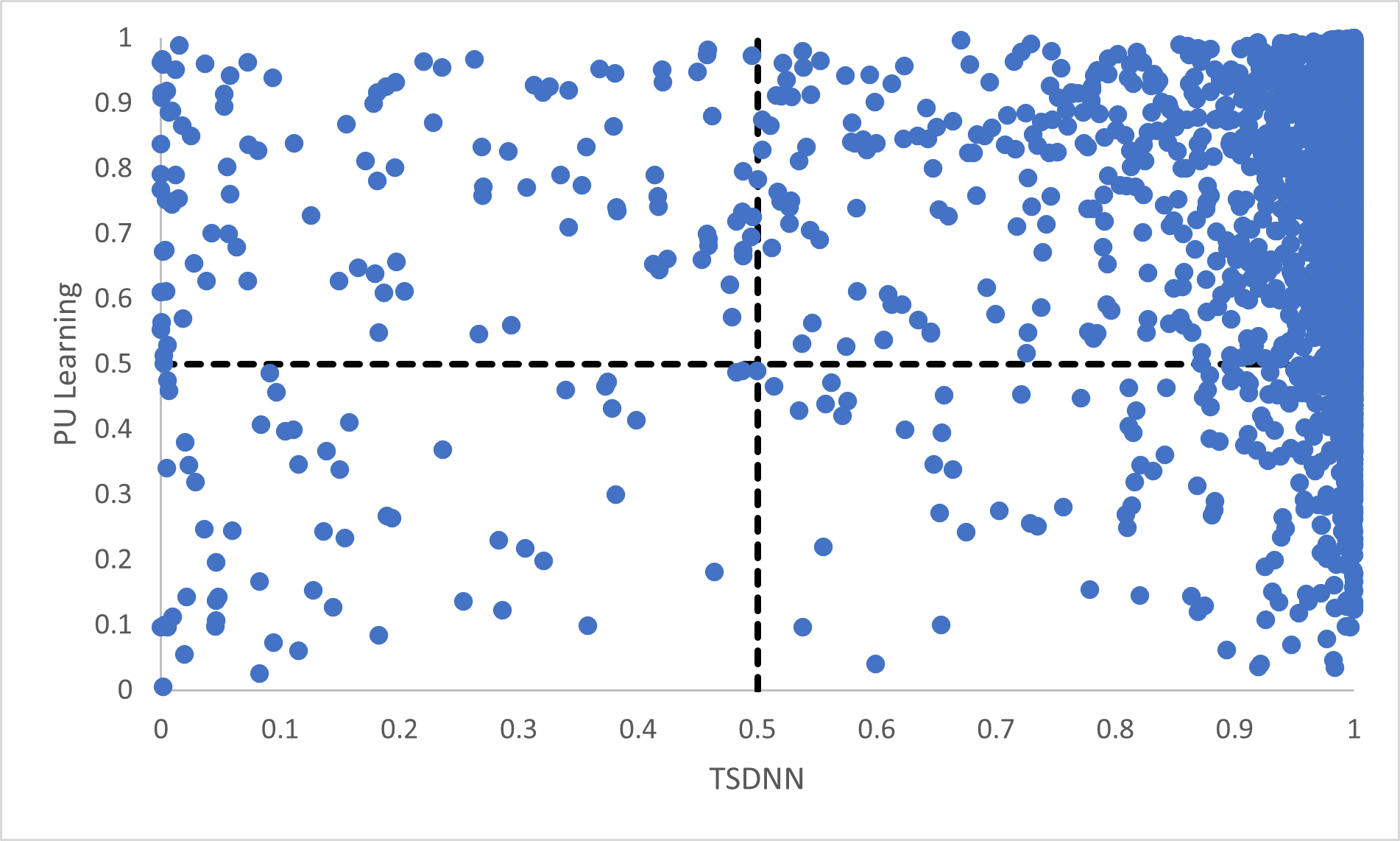}
  \caption{Scatter plot of our TSDNN predicted scores vs PU learning on ICSD materials from the test set. This shows that the PU learning method falsely classified many more materials as negative (Quadrant IV) than our TSDNN model (Quadrant II) for the PU learning predictions.}
  \label{fig:TsdnnPuComparison}
\end{figure}

In both the basic PU learning method for synthesizability \cite{jang2020structure} and our TSDNN framework, a decision boundary of 0.5 is used for determining synthesizable vs. unsynthesizable materials for both classifiers. To show the consistency and performance of both models, Figure \ref{fig:TsdnnPuComparison} plots the probabilities of being stable materials for all the ICSD materials from our test set by the PU learning model against those predicted by our TSDNN model. The figure is divided into quadrants, with each quandrant signifying agreement or disagreement between the PU learning method and our TSDNN framework. The top right quadrant signifies correct agreement between the models, where both models correctly classify the materials as positive. The bottom left quadrant, similarly, denotes the incorrect agreement that the materials should be classified as negative. The bottom right quadrant signifies a disagreement in which the TSDNN model correctly classifies the materials and the PU learning method does not. It can be easily found that the bottom right quadrant contains much more samples compared to the top left quadrant,  solidly indicating that there are many materials with very high prediction scores correctly predicted by our TSDNN model, but were incorrectly classified by the PU learning method as being unstable (high formation energy). There were comparatively few materials in the top left quadrant that are samples correctly classified by the PU learning method and incorrectly classified by the TSDNN model. These results show that while our model has improved true positive rate, the improvement is not simply a result of materials being classified right at the 0.5 boundary.

\subsubsection{Formation energy classification performance} \label{sec:fe_perfomance}

Formation energy based materials screening can be done using either regression models or classification models, depending on the motivation of the screening. For screening hypothetical materials, the first step is identifying potentially stable candidates with negative formation energies. As the exact formation energy is not needed, this can be done effectively by an accurate formation energy classification model. To evaluate the performance of models for formation energy classification, we consider accuracy, precision, and F1 score, as each metric corresponds to a specific screening motivation. We notably do not use recall as for our problem here, simply achieving a high recall may not be meaningful on its own because it may include many false positives that are not stable. F1 score better represents performance in this regard, as it measures the performance with balanced recall and precision. In this situation, predicting few false-positives while still correctly classifying a majority of the actual positive materials is desired. For precision, in situations in which it is imperative that the screened materials be below a given eV threshold (e.g. finding materials with high-confidence stability), a high-precision model is the most optimal choice regardless of its accuracy or F1 score. Precision and F1 score are useful metrics at any eV threshold. Accuracy, however, is only significant with an eV threshold of 0.0 eV for our test set as we are seeking to classify between samples with negative or positive formation energies. With lower eV thresholds, the number of negative samples vastly outweighs the number of positive samples, as shown in Figure \ref{fig:cubicgan_fe_hist}. A model could have a high accuracy at a low eV threshold while correctly classifying few actual positive samples. Accuracy is most useful for identifying >0.0 eV/atom materials. A high-accuracy model with a threshold of 0.0 eV achieves the best balance between correctly identifying actual positive and negative samples.

\begin{figure}[ht]
\centering
    \begin{subfigure}[h]{0.4\textwidth}
        \includegraphics[width=\textwidth]{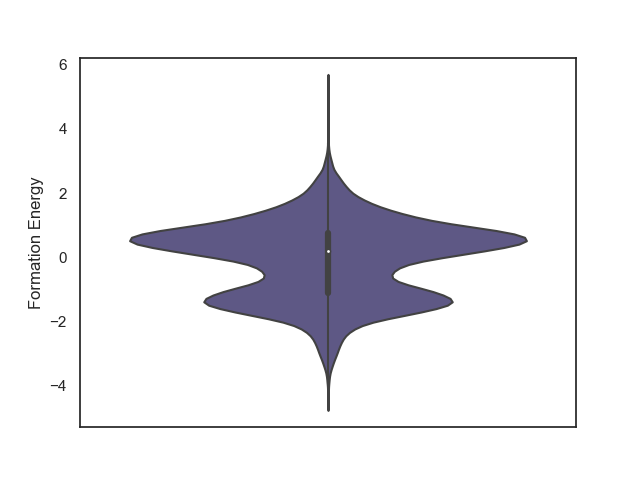}
        \caption{Ground truth Ef distribution of test set}
        \vspace{-3pt}
        \label{fig:all_fe_violin}
    \end{subfigure}
    \begin{subfigure}[h]{0.4\textwidth}
        \includegraphics[width=\textwidth]{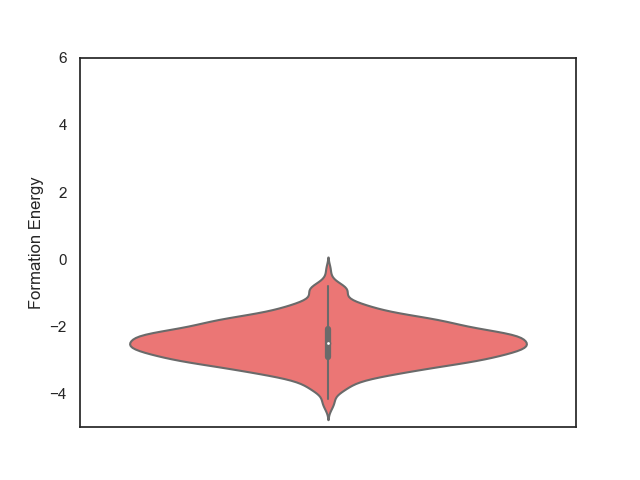}
        \caption{Ef predictions by Separated TSDNN}
        \vspace{-3pt}
        \label{fig:low_fe_violin}
    \end{subfigure}
    \begin{subfigure}[h]{0.4\textwidth}
        \includegraphics[width=\textwidth]{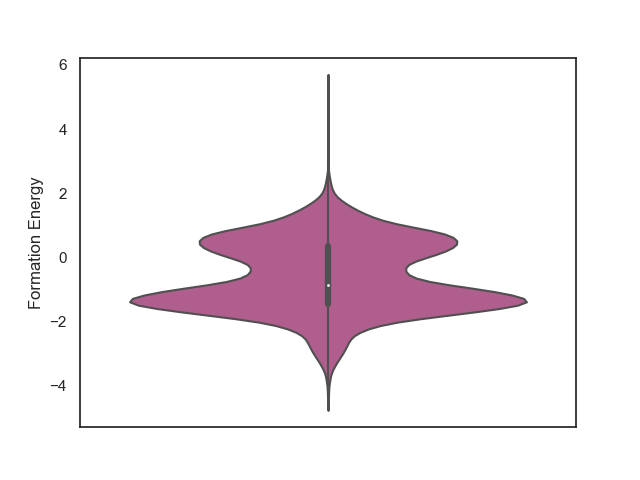}
        \caption{Ef predictions by Unseparated TSDNN }
        \vspace{-3pt}
        \label{fig:distributed_violin}
    \end{subfigure}
    \begin{subfigure}[h]{0.4\textwidth}
        \includegraphics[width=\textwidth]{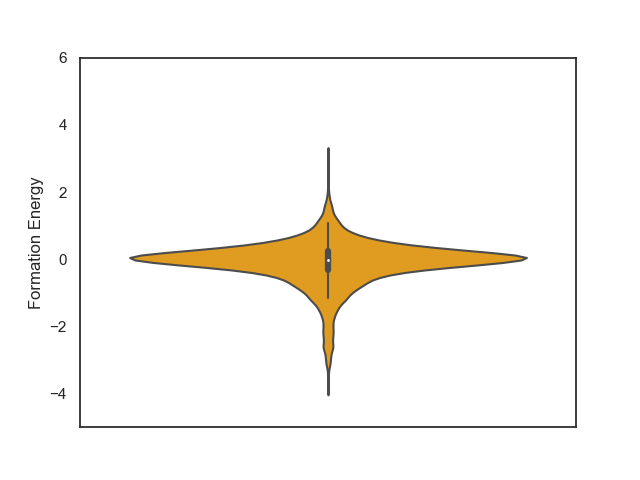}
        \caption{Ef predictions by CGCNN}
        \vspace{-3pt}
        \label{fig:cgcnn_violin}
    \end{subfigure}
    \caption{Comparison of the formation energy distributions of test samples predicted/classified to have negative formation energy by three different models versus the ground truth. (a) The distribution of the formation energies of all test samples. 35\% of them are positive. (b) Distribution of Ef of positive samples predicted by the Separated TSDNN model. (c) Distribution of Ef of positive samples predicted by the Unseparated TSDNN model. (d) Ef distribution of positive samples predicted by the CGCNN regression model. }
    \label{fig:fe_distributions}
\end{figure}

\begin{table}[h]
\centering
\caption{Comparison of classification performance for formation energy with an eV threshold of 0.0.}
\begin{tabular}{|l|l|l|l|}
\hline
\multicolumn{1}{|c|}{Model} & \multicolumn{1}{c|}{Precision} & \multicolumn{1}{c|}{F1 Score} & \multicolumn{1}{c|}{Accuracy} \\ \hline
CGCNN       & 58.60\%     & 64.30\%     & 64.30\%             \\ \hline
Unseparated FE TSDNN   & 66.10\%   & \textbf{74.60\%}   & \textbf{74.00\%}        \\ \hline
Separated FE TSDNN      & 100.00\%  & 16.50\%   & 59.50\%       \\ \hline

\end{tabular}
\label{tab:zero_fe_performance}
\end{table}

Table \ref{tab:zero_fe_performance} shows the classification performance of three models on our test set of materials. 
Our Unseperated TSDNN model achieves a 74.60\% F1 score compared to the CGCNN regression model's F1 score of 64.3\%, with a significant absolute 10.3\% improvement by using our semi-supervised learning approach. At the same time, this model achieves an accuracy of 74\% , with an absolute 9.7\% improvement over the CGCNN model. Our Separated TSDNN model shows that our approach is able to be tweaked for achieving higher precision by adjusting the training threshold, resulting in a high-confidence model. Here the table shows that the model can be tuned to achieve 100\% precision for identifying candidates which are highly likely to be stable materials.

To further illustrate the advantage of our TSDNN models, we show the formation energy distributions of the positively classified samples (with negative formation energies) from our test set by our classifiers and the baseline CGCNN regression model. As shown in Figure \ref{fig:all_fe_violin}, our test set contains a large number of samples with positive formation energy to fully test the model's ability to differentiate between samples with positive and negative formation energy. The desired formation energy distribution of screened samples is seen in the bottom group of samples around -2.0 eV. Figure \ref{fig:low_fe_violin} shows that our Seperated TSDNN model has just obtained the desired sample groups with the formation energy distributed around the peak of -2.2eV, which indicates that our Separated FE TSDNN is effective for applications which require a high certainty that a material will have a low formation energy because of its very high precision. For more general screening with an eV threshold of 0.0, our Unseparated TSDNN model is more suitable (Figure \ref{fig:distributed_violin}). With the vast array of materials with formation energies very close to 0.0 eV, it is very challenging to train a model to accurately differentiate between materials with small positive and small negative formation energy. As shown in Figure \ref{fig:cgcnn_violin}, the CGCNN model is not able to capture the full distribution of negative formation energy materials in the test set and has difficulty in differentiating between samples with positive and negative formation energies. As shown in Table \ref{tab:zero_fe_performance}, our Unseparated TSDNN model is able to improve greatly in performance with a 7.5\% increase in precision, a 10.3\% increase in F1 Score, and a 9.7\% increase in accuracy from the CGCNN. This makes it preferred for applications that wish to screen for stable materials (usually with negative formation energy).


\subsection{New materials discovery using both formation energy and synthesizability screening models}

\subsubsection{Generation of candidate cubic structures for screening}


CubicGAN~\cite{zhao2021high} is a generative adversarial network based model for generating novel cubic crystal structures. CubicGAN reports that when generating 10 million virtual cubic crystal structures, most of materials in training datasets, Materials Project and ICSD can rediscovered. Thus, We use CubicGAN to generate 10 millions of virtual cubic crystal structures, of which around 90\% materials can be recognized as the same space groups they are assigned to. The next step is to remove duplicate crystal structures. We consider materials with the same compositions and the same corresponding atom positions as duplicate materials. Around 25\% materials (~2.5 millions) are left for further analysis.

Starting with 2.5 millions of candidate materials, we first apply our Separated TSDNN model to classify them into positive or negative formation energies. 918686 of them are predicted as having a negative formation energy. We then select 5000 of these materials with the highest prediction scores and apply our Unbalanced TSDNN synthesizability model to predict their probability of being able to be synthesized. We finally select the top 1000 samples with the highest probability to be synthesizable. These samples are sent for DFT relaxation and further validation.

\subsubsection{DFT validation of predicted candidate structures}
The density functional theory (DFT) based first principle calculations were performed using the  Vienna \textit{ab initio} simulation package (VASP) \cite{Vasp1,Vasp2,Vasp3,Vasp4}. The electron-ion interactions were treated employing the projected augmented wave (PAW) pseudopotentials where 520 eV plane-wave cutoff energy was set\cite{PAW1, PAW2}.  The exchange-correlation functional was considered with the generalized gradient approximation (GGA) based on the Perdew-Burke-Ernzerhof (PBE) method \cite{GGA1, GGA2}. The energy convergence criterion was set as 10$^{-5}$ eV, while the atomic positions were optimized with the force convergence criterion of 10$^{-2}$ eV/{\AA}. The Brillouin zone integration for the unit cells was computed using the $\Gamma$-centered  Monkhorst-Pack $k$-meshes. The Formation energies (in eV/atom) of several materials were determined based on the expression in  Eq.~\ref{eq:form}, where $E[\mathrm{Material}]$ is the total energy per unit formula of the considered structure, $E[\textrm{A}_i]$ is the energy of $i^\mathrm{th}$ element of the material, $x_i$ indicates the number of A$_i$ atoms in a unit formula, and $n$ is the total number of atoms in a unit formula($n=\sum_i x_i$). 

\begin{equation}
    E_{\mathrm{form}} =\frac{1}{n}(E[\mathrm{Material}] - \sum_i x_i E[\textrm{A}_i])
    \label{eq:form}
\end{equation}

\begin{figure}[h]
\centering
\includegraphics[width=0.6\textwidth]{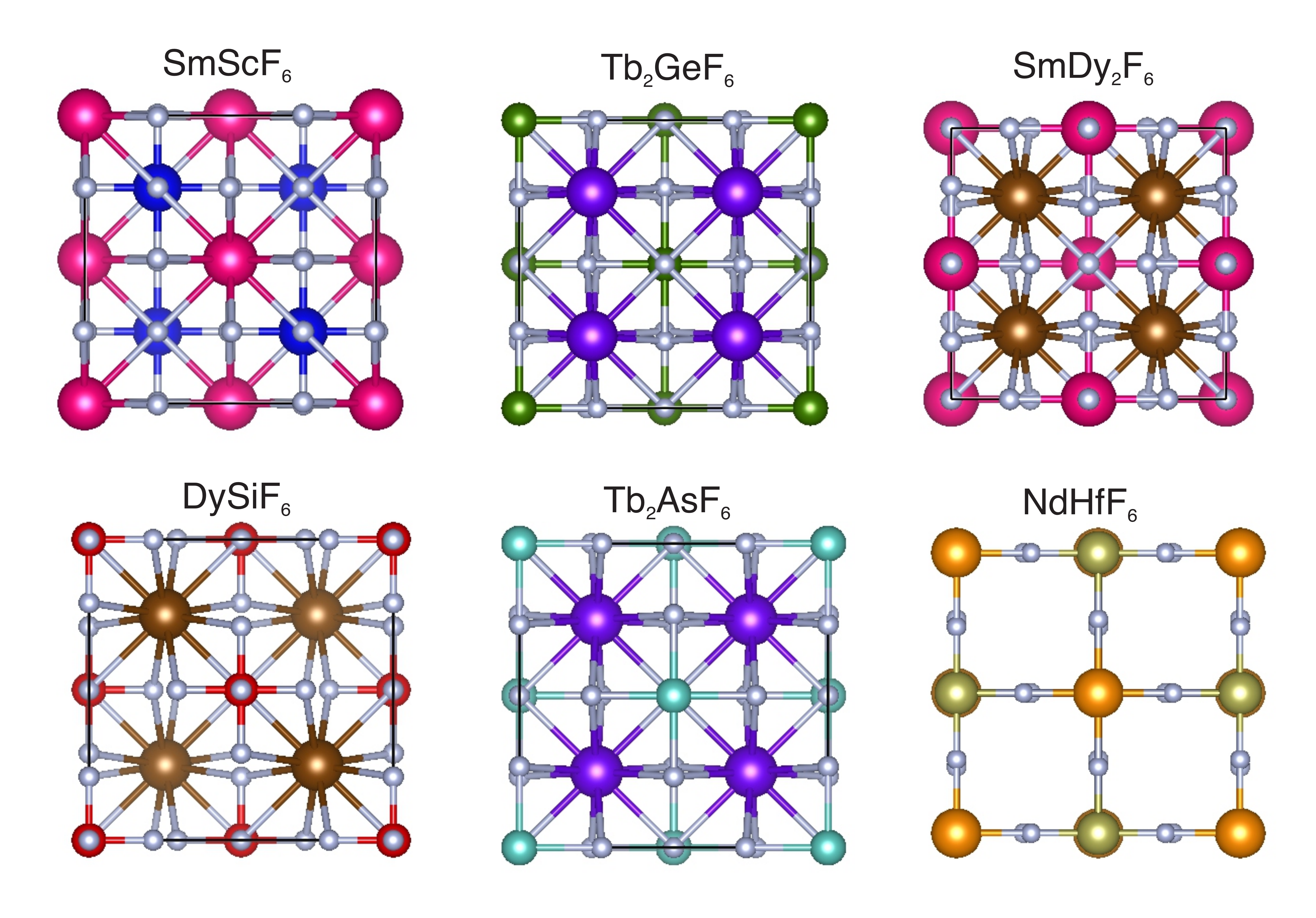}
\caption{The discovered new crystal structures with lowest formation energies.}
\label{fig:structures}
\end{figure}

Out of 1000 crystal structures, which were optimized using DFT, 512 of them have negative formation energies. Table~\ref{tab:DFT} shows the 10 cubic structures found with lowest formation energies. Interestingly, all the 10 materials have rare-earth elements. Half of them have ADF$_6$ type chemical formulas and the other half have A$_2$DF$_6$ formulas. In all these structures, F is the common element, and the rest of the elements make bonds with F (see Fig.\ref{fig:structures}). 

\begin{table}[h]
\centering
\caption{ The chemical formulas and the space group symmetries for the materials found with lowest formation energies. }
\label{tab:DFT}
\begin{tabular}{|c|c|c|}
\hline
Chemical Formula & Space Group Number & E$_\mathrm{form}$ (eV/atom)  \\ \hline
SmScF6           & 216                & -4.1849 \\ \hline
Tb2GeF6          & 225                & -4.0848 \\ \hline
SmDy2F6          & 225                & -3.9257 \\ \hline
DySiF6           & 216                & -3.8332 \\ \hline
Tb2AsF6          & 225                & -3.8213 \\ \hline
NdHfF6           & 225                & -3.7355 \\ \hline
Dy2TaF6           & 225                & -3.6995 \\ \hline
LaHoF6           & 216                & -3.5676 \\ \hline
TmScF6           & 216                & -3.5272 \\ \hline
Nd2GaF6           & 216                & -3.4555 \\ \hline
\end{tabular}
\end{table}



\section{Discussion}
\label{sec:others}

With the advent of large-scale material databases and  generative machine learning models, an immense expanse of the wider inorganic material chemical design space is now possible with high throughput experiments or computation. This extensive amount of data makes it a prime target for developing machine learning models for both synthesizability and formation energy based screening. However, there is comparatively little labeled data in both cases and particularly few negative samples. Obtaining new labeled data can be both costly, time-consuming, and unreliable. 

\begin{figure}[ht]
  \centering
  \includegraphics[width=0.6\linewidth]{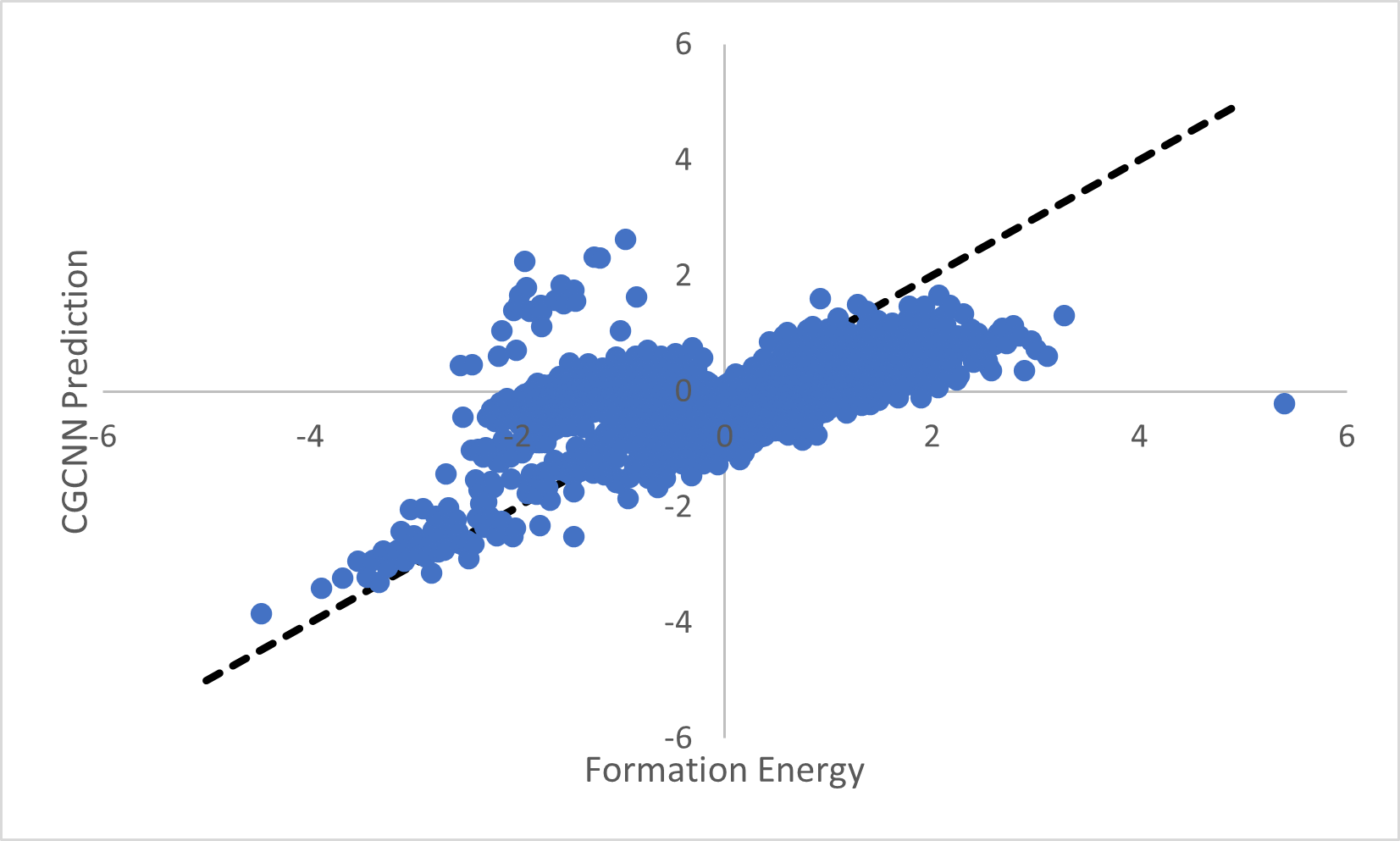}
  \caption{Scatter plot of CGCNN predicted formation energy. With few samples with positive formation energy, the CGCNN model tends to underestimate true positive formation energy materials and overestimate true negative formation energy materials. Furthermore, it seems that the CGCNN has greatly overestimated a portion of the materials substantially.}
  \label{fig:cgcnn_regplot}
\end{figure}

Previously, CGCNN-based regression models have been used to screen for stable material candidates using predicted formation energy. The issue with such models to screen for materials candidates with low formation energies is the introduction of model and prediction biases due to the dataset imbalance. As shown in Figure \ref{fig:mp_fe_hist}, only 8.2\% of the total MP database is comprised of materials with formation energy greater than 0 eV. This results in ML based regression models that bias their predictions heavily toward negative formation energies with true positive samples, as shown in Figure \ref{fig:cgcnn_regplot}. 

Here we proposed a dual crystal graph convolutional neural network-based semi-supervised learning framework for synthesizability and formation energy prediction. Comprehensive testing and validation show that our TSDNN models can successfully exploit the unlabeled data in each use case in conjunction with existing labeled data to accurately and effectively predict synthesizability and formation energies. Our TSDNN models can be paired with existing and future material generation models for efficient screening across a variety of applications, as shown with our CubicGAN. Our models’ integration with generative models provides for a greatly optimized and more reliable search for new materials. Compared to the CGCNN based regression model, which misclassified a large grouping of materials as having positive formation energies due to the bias caused by the dataset imbalance, our semi-supervised TSDNN classification model reduces this bias, as it is designed with screening in mind from start. Furthermore, by using our TSDNN framework in conjunction with our CubicGAN model, we were able to use the large amount of unscreened data as unlabeled data to train our model for improved performance.


Another major finding of our study is that the evaluation method plays an important role in for objective evaluation of model performance. In previous studies \cite{jang2020structure}, the model performance has been evaluated using a standard random holdout validation, but with only the positive class. While reliable for other problem domains, this is inadequate for inorganic material classification as it allows for possible too high compositional and structural similarity between the test and training data to artificially inflate classification performance. This is due to the tinkering materials discovery process over history, which leads to that materials deposited in materials repositories such as Materials Project are grouped into clusters with high similarity, which can lead to over-estimation of materials property prediction models \cite{meredig2018can,xiong2020evaluating}. Without a negative class in the test set, there is no guarantee that the model simply predicts every sample as positive indescriminantly. For instance, when evaluated using a random holdout test set, our synthesizability prediction model was able to achieve a true positive rate of 97.90\%. It is for this reason that we elected to introduce a negative class composed of the lowest classification score materials, as shown in Figure \ref{fig:pu_learning_framework}.


\section{Conclusion}
\label{sec:others}
Machine learning based materials property prediction faces the big challenge of lack of sufficient annotated property data and the issue of missing negative samples (non-stable materials), which is needed for building screening models for new materials discovery. 
To address these two issues, we propose a teacher-student twin graph neural network model (TSDNN) for materials property prediction using formation energy and synthesizability as examples. We formulate both problems as a semi-supervised binary classification problem which matches well to the real-world screening scenarios where these ML screening models are used to pick stable and synthesizable materials candidates from the big pool of hypothetical materials. Our extensive experiments show that our TSDNN models are able to significantly improve the prediction performance compared to previous methods in both synthesizability and formation energy prediction. We achieve a 92.9\% true positive rate for synthesizability prediction with a much simpler model architecture and 74\% prediction accuracy for formation energy screening. As further validation, we applied our models to the 2,545,713 hypothetical materials generated by our CubicGAN model. Overall, we screened 918686 materials that were positively classified by the formation energy model with our synthesizability prediction model. We select the top 1000 of these final screened materials for DFT verification and find that 51.2\% have negative formation energies. These results show that our TSDNN semi-supervised learning framework is effective for large-scale material discovery screening.

\section{Availability of data and code}

The data that support the findings of this study are openly available in Materials Project database at \href{http:\\www.materialsproject.org}{\textcolor{blue}{http:\\www.materialsproject.org}}. The source code and our pretrained models are  freely available at our github repository \href{https://github.com/usccolumbia/tsdnn}{\textcolor{blue}{https://github.com/usccolumbia/tsdnn}}


\section{Contribution}
Conceptualization, J.H.; methodology, D.G., J.H., E.S, Y.Z., N.F.; software, D.G.; validation, E.S., J.H.;  investigation, J.H., D.G., E.S., Y.Z.; resources, J.H.; data curation, J.H., and Y.Z.; writing--original draft preparation, D.G., J.H., E.S. ; writing--review and editing, J.H, D.G., N.F.; visualization, D.G.; supervision, J.H.;  funding acquisition, J.H.

\section{Declaration of conflict of interests}

The authors declare there no conflict of interests.

\section{Acknowledgement}
Research reported in this work was supported in part by NSF under grants 1940099 and 1905775. The views, perspective, and content do not necessarily represent the official views of NSF. This work was supported in part by the South Carolina Honors College Research Program. This work is partially supported by a grant from the University of South Carolina Magellan Scholar Program.

\bibliographystyle{unsrt}  
\bibliography{references}  
\end{document}